\documentclass[%
 aip,
 rsi,
amsmath,amssymb,
reprint,
floatfix
]{revtex4-2}

\usepackage{graphicx}
\usepackage{dcolumn}
\usepackage{bm}

\usepackage[utf8]{inputenc}
\usepackage{mathptmx}
\usepackage{braket}
\usepackage[english]{babel}
\usepackage{xcolor}
\colorlet{RED}{red}
\colorlet{BLUE}{blue}
\usepackage{dcolumn}
\usepackage{bm}
\usepackage[version=4]{mhchem} 
\usepackage{acronym}
\usepackage{adjustbox}
\usepackage{tikz}
\usepackage[toc,page]{appendix}
\usepackage{microtype} 
\usetikzlibrary{calc,shapes.geometric,decorations.pathmorphing,patterns}

\definecolor{background-color}{gray}{0.98}
\usepackage[margin=2.3cm,bmargin=1cm,footnotesep=1cm]{geometry}

\newcommand{\PNNL}{%
   \affiliation{%
       Physical Sciences Division, Pacific Northwest National Laboratory, Richland, WA 99354, USA
       }}

\begin{document}

\title{
%
Sub-system self-consistency in coupled cluster theory
}

\author{Karol Kowalski} 
\email{karol.kowalski@pnnl.gov} \PNNL 

\date{\today}

\begin{abstract}
In this Communication, we provide numerical evidence indicating that the single-reference coupled-cluster (CC) energies can be calculated alternatively to its copybook definition. We demonstrate that the CC energy can be reconstructed by diagonalizing the effective Hamiltonians describing correlated sub-systems of the many-body system.
In the extreme case, we provide numerical evidences that the CC energy can be reproduced through the diagonalization of the effective Hamiltonian describing sub-system composed of a single electron. 
These properties of the CC formalism can be exploited to design protocols to 
define effective interactions in sub-systems used as  probes to calculate the energy of the entire system and introduce a new type of self-consistency for approximate CC approaches. 
\end{abstract}

\maketitle

%
%
%
%
%

\section{Introduction}
The standard single-reference coupled cluster (SR-CC) theory 
\cite{coester58_421,coester60_477,cizek66_4256,paldus72_50,purvis82_1910,arponen83_311,bishop1987coupled,paldus07,crawford2000introduction,bartlett_rmp}
is a direct consequence of the linked cluster theorem (LCT)
and represents  the simplest case of exponential Ansatz for the ground-state wave functions. 
Various SR-CC formulations have been widely used in many  areas of physics and chemistry  to describe systems and processes driven by complex correlation effects.
\cite{scheiner1987analytic,sinnokrot2002estimates,slipchenko2002singlet,tajti2004heat,crawford2006ab,parkhill2009perfect,riplinger2013efficient,yuwono2020quantum,stoll1992correlation,hirata2004coupled,katagiri2005equation,booth2013towards,degroote2016polynomial,mcclain2017gaussian,wang2020excitons,PhysRevX.10.041043}

Recently, the LCT has been generalized to the active spaces formulations using sub-system embedding sub-algebras CC formalism (SES-CC),\cite{safkk,kowalski2021dimensionality} which led to new alternative ways for calculating  CC energy and allowing one to interpret CC theory as a renormalization (downfolding) procedure.  
From this point of view, the SR-CC theory can be viewed as a renormalization procedure or rigorous embedding algorithm. The SES-CC formalism or closely related double unitary CC formulations  have already inspired several developments to describe strongly correlated systems  \cite{bauman2019downfolding,metcalf2020resource,bauman2022coupled} and time-evolution of quantum systems.\cite{downfolding2020t}
(see also Refs.\onlinecite{jankowski1996approximate,nooijen1999combining,he2022second,callahan2021dynamical,kvaal2022three} for related developments).

In this paper, for the first time in the literature, we provide numerical evidence that the coupled cluster energy for a given molecular basis can be calculated alternatively to the textbook energy formula. Using sub-system embedding  SES-CC formalism, the same energy can be obtained by diagonalizing effective Hamiltonians in appropriate active spaces. These results can be utilized to design a new class of CC approximations, including those that provide a new way of encapsulating the sparsity (locality)  of the quantum system. The goal of this paper is not to discuss the accuracies of the particular CC approximation but rather to illustrate general and newly discovered properties of the SR-CC formulations stemming from the SES-CC formalism. In particular, we 
provide a numerical illustration of the so-called SES-CC theorem, which states that the CC energies can be alternatively obtained by diagonalization of the appropriately defined effective Hamiltonians and 
answer the question: how many ways can the energy of approximate or exact CC formulations be calculated for a fixed orbital basis set? 
In this Communication, we also demonstrate that the sub-systems wave functions ({\it vide infra}) may also break the spin-symmetry of the composite system. This is the case when the definition of the active spaces employing active orbitals used by the SES-CC theorem is extended to active spaces defined by spin-orbitals. In such situations, effective Hamiltonians can also break the spin symmetry of the original Hamiltonian describing the whole system. We provide a numerical illustration of the validity of the SES-CC theorem in both cases. It is compelling to notice that the SES-CC formalism can reproduce the CC energy even by using a sub-system composed of one "dressed" electron.
To illustrate the above-mentioned properties of the SES-CC formalism, we use several benchmarks such as the H4, H6, H8, and Li$_2$ systems. For the H4 and H6 models, we consider two geometrical configurations corresponding to weakly and strongly correlated regimes of the ground-state wave functions. 

\section{SES-CC formulation} 
The SR-CC formulation is defined through  the exponential Ansatz for the ground-state wave function 
$|\Psi\rangle$, 
\begin{equation}
|\Psi\rangle = e^T |\Phi\rangle \;,
\label{cc1}
\end{equation}
where $T$ and $|\Phi\rangle$ represent the so-called cluster operator and single-determinantal reference function. The cluster operator is defined  by its many-body components 
$T_k$
\begin{equation}
    T = \sum_{k=1}^M T_k =  \sum_{k=1}^M \frac{1}{(k!)^2} \sum_{i_1,\ldots,i_k; a_1,\ldots, a_k} t^{i_1\ldots i_k}_{a_1\ldots a_k} E^{a_1\ldots a_k}_{i_1\ldots i_k} \;.
    \label{cc2}
\end{equation}
In the above equation indices $i_1,i_2,\ldots$ ($a_1,a_2,\ldots$) refer to occupied (unoccupied) spin-orbitals in the reference function $|\Phi\rangle$.
The excitation operators $E^{a_1\ldots a_k}_{i_1\ldots i_k} $ are defined through strings of standard creation ($a_p^{\dagger}$) and annihilation ($a_p$)
operators
\begin{equation}
E^{a_1\ldots a_k}_{i_1\ldots i_k}  = a_{a_1}^{\dagger}\ldots a_{a_k}^{\dagger} a_{i_k}\ldots a_{i_1} \;.
\label{estring}
\end{equation}
When $M$ in the summation in Eq. (\ref{cc2}) is equal to the number of correlated electrons ($N$), then the corresponding CC formalism is equivalent to the full configuration interaction (FCI) method, otherwise for $M<N$ one deals with the standard approximation schemes. Typical CC formulations such as CCSD, CCSDT, and CCSDTQ correspond to $M=2$, $M=3$, and $M=4$ cases, respectively.
\cite{purvis82_1910,ccsdt_noga,ccsdt_noga_err,scuseria_ccsdt,ccsdtq_nevin,Kucharski1991}

The equations  
for cluster amplitudes $t^{i_1\ldots i_k}_{a_1\ldots a_k}$ and ground-state energy $E$ can be obtained by introducing Ansatz (\ref{cc1}) into the Schr\"odinger equation and projecting onto $P+Q$ space, where $P$ and $Q$ are the projection operator onto the reference function and the space of excited Slater determinants obtained by acting with the cluster operator onto the reference function $|\Phi\rangle$, i.e., 
\begin{equation}
    (P+Q) H e^T |\Phi\rangle = E (P+Q) e^T |\Phi\rangle \;,
    \label{sch1}
\end{equation}
where $H$ represents the electronic Hamiltonian. The above equation is the so-called energy-dependent form of the CC equations, which is equivalent to the  eigenvalue problem only in the exact wave function limit 
when $T$ contains all possible excitations. However, the above equations for approximate CC formulations do not represent the standard eigenvalue problem. At the solution, the energy-dependent CC equations are equivalent to the 
energy-independent  equations:
\begin{eqnarray}
Qe^{-T}He^T |\Phi\rangle &=& 0 \;, \label{coeq1} \\
 \langle\Phi|e^{-T} H e^{T} |\Phi\rangle &=& E\;. \label{coeq2}
\end{eqnarray}
where only  connected diagrams contribute to Eqs.(\ref{coeq1}) and (\ref{coeq2}). 


In the following part of the discussion,  we will focus on the closed-shell CC formulations that use the restricted Hartree-Fock (RHF) Slater determinant as a reference function. 
The main idea of SES-CC formulation hinges upon the 
characterization of sub-systems of a quantum system of interest in terms of active spaces or commutative sub-algebras of excitations that define corresponding active space. 
To this end we  introduce  sub-algebras of algebra $\mathfrak{g}^{(N)}$ generated by   
$E^{a_l}_{i_l}=a_{a_l}^{\dagger} a_{i_l}$ operators in the particle-hole  representation defined  with respect to the reference $|\Phi\rangle$. As a consequence of using the particle-hole formalism,  all generators 
commute, i.e., $[E^{a_l}_{i_l},E^{a_k}_{i_k}]=0$, and algebra $\mathfrak{g}^{(N)}$  (along with all sub-algebras considered here) is commutative.
The SES-CC approach employs  class of sub-algebras  of commutative
$\mathfrak{g}^{(N)}$ algebra,  which contain all possible excitations 
$E^{a_1\ldots a_m}_{i_1\ldots i_m}$ needed to generate  all possible excitations from a subset of active occupied orbitals (denoted as $R$, $\lbrace R_i, \; i=1,\ldots,x_R \rbrace$)
to a subset of active virtual orbitals (denoted as $S$, $\lbrace S_i, \; i=1,\ldots,y_S \rbrace$) defining active space. 
These sub-algebras will be designated as $\mathfrak{g}^{(N)}(R,S)$.
Sometimes it is convenient to use alternative notation
$\mathfrak{g}^{(N)}(x_R,y_S)$ where numbers of active orbitals in $R$ and $S$ orbital  sets, $x_R$ and $y_S$, respectively,  are explicitly called out.
As discussed in Ref.\onlinecite{safkk}, configurations  generated by elements of $\mathfrak{g}^{(N)}(x_R,y_S)$,  along with the reference function,
span the complete active space (CAS) referenced to as the CAS($R,S$)
(or equivalently CAS($\mathfrak{g}^{(N)}(x_R,y_S)$)).
In the same way, one can define sub-algebras defined by a chosen subsets of spin-orbitals.
%
%
%
%
%


In  previous papers on this topic (see Refs. \onlinecite{safkk,downfolding2020t,kowalski2021dimensionality}), we analyzed the  effect of the partitioning of the cluster operator induced by general  sub-algebra $\mathfrak{h}=\mathfrak{g}^{(N)}(x_R,y_S)$, where the cluster operator $T$, given by Eq. (\ref{cc2}), is represented as
\begin{equation}
T=T_{\rm int}(\mathfrak{h})+T_{\rm ext}(\mathfrak{h}) \;,
\label{deco1}
\end{equation}
where $T_{\rm int}(\mathfrak{h})$ belongs to $\mathfrak{h}$ while 
$T_{\rm ext}(\mathfrak{h})$ does no belong to $\mathfrak{h}$.
If the expansion $T_{\rm int}|\Phi\rangle$ produces all Slater determinants (of the same symmetry as the $|\Phi\rangle$ state) in the active space, we call $\mathfrak{h}$ the {\it sub-system embedding sub-algebra} for the CC formulation defined by the $T$ operator. 
\begin{figure}
    \centering
    \includegraphics[width=\linewidth]{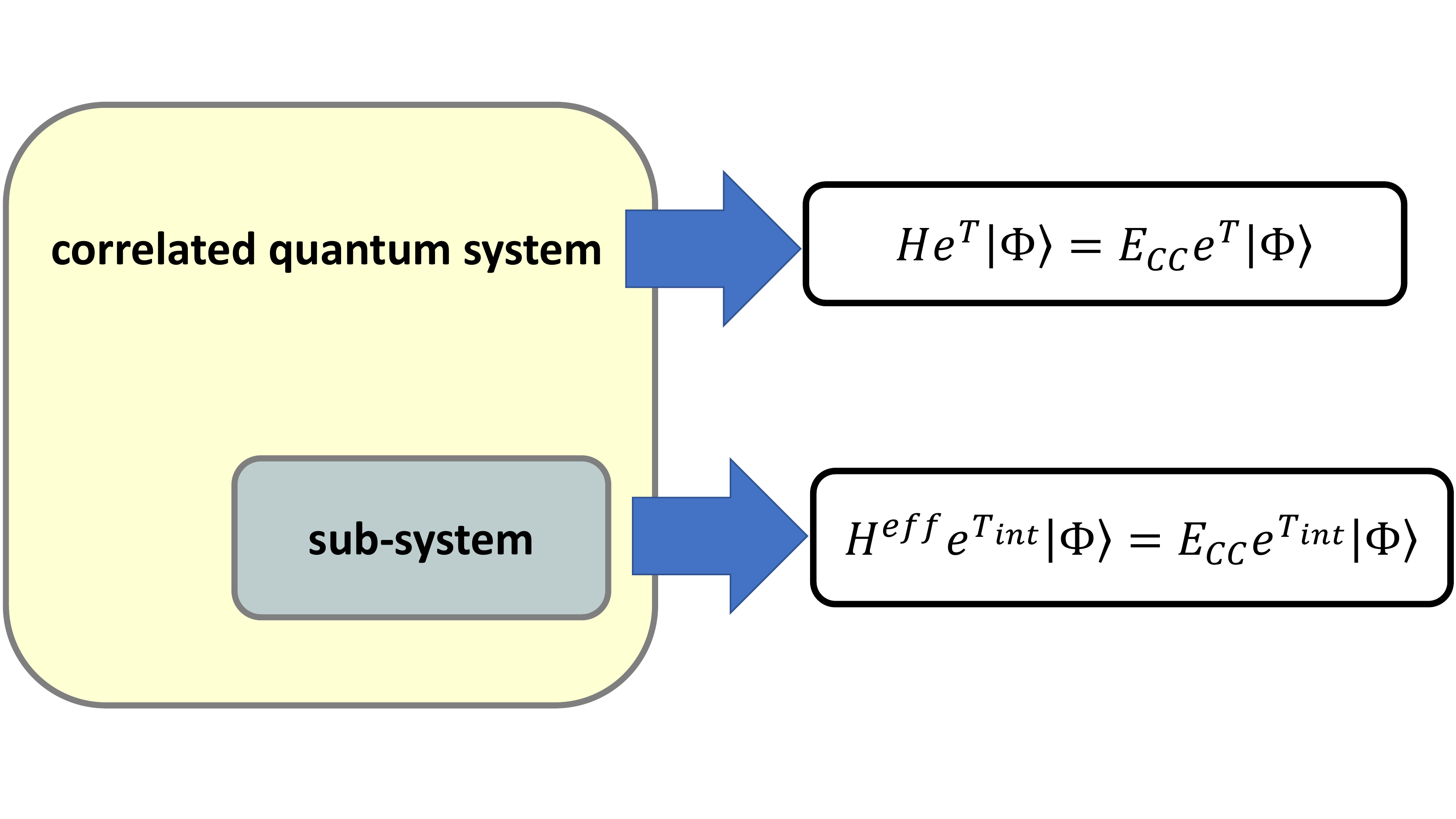}
    \caption{Schematic representation of the SES-CC theorem. The CC energy can be obtained as an effective Hamiltonian eigenvalue.}
    \label{fig1}
\end{figure}
In Ref.\onlinecite{safkk}, we showed that CC approximations have specific  classes of SESs. 
A  consequence of  the existence of SESs for standard CC approximations is the fact that the corresponding energy can be calculated, in an alternative way to Eq. (\ref{coeq2}), as an eigenvalue of the active-space non-Hermitian eigenproblem (this fact will be referred to as the {\it SES-CC theorem} (see Fig.\ref{fig1}))
 \begin{equation}
    H^{\rm eff}(\mathfrak{h})
    e^{T_{\rm int}(\mathfrak{h})}|\Phi\rangle = 
    E e^{T_{\rm int}(\mathfrak{h})}|\Phi\rangle  \;.
\label{seseqh}
\end{equation}
where 
\begin{equation}
H^{\rm eff}(\mathfrak{h})=(P+Q_{\rm int}(\mathfrak{h})) \bar{H}_{\rm ext}(\mathfrak{h}) (P+Q_{\rm int}(\mathfrak{h}))\;
\label{heffses}
\end{equation}
and 
\begin{equation}
\bar{H}_{\rm ext}(\mathfrak{h})=e^{-T_{\rm ext}(\mathfrak{h})} H e^{T_{\rm ext}(\mathfrak{h})} \;.
\label{heffdef}
\end{equation}
In Eq.(\ref{heffses}), the projection operator $Q_{\rm int}(\mathfrak{h})$
is a projection operator on a sub-space spanned by  all Slater determinants generated by $T_{\rm int}(\mathfrak{h})$ acting onto $|\Phi\rangle$. 
The wave function $|\Psi(\mathfrak{h})\rangle$ defined as
\begin{equation}
|\Psi(\mathfrak{h})\rangle = e^{T_{\rm int}(\mathfrak{h})}|\Phi\rangle \;,
\label{subwf}
\end{equation}
correlates electrons within active space (CAS($\mathfrak{h}$)) while leaving the remaining part of the system uncorrelated. For this reason, we call
$|\Psi(\mathfrak{h})\rangle$ a {\it  sub-system wave function}. These results can be easily extended to the active spaces defined at the spin-orbital level. However, for the closed-shell reference 
function the use of active spaces defined at the level of spin-orbitals  leads to the sub-system wave functions $|\Psi(\mathfrak{h})\rangle$ and corresponding effective Hamiltonian 
$H^{\rm eff}(\mathfrak{h})$ that break the symmetry of the reference function $|\Phi\rangle$ and total Hamiltonian $H$, respectively (see Fig.\ref{fig:siam_results}).
\begin{figure}
    \centering
    \includegraphics[width=\linewidth]{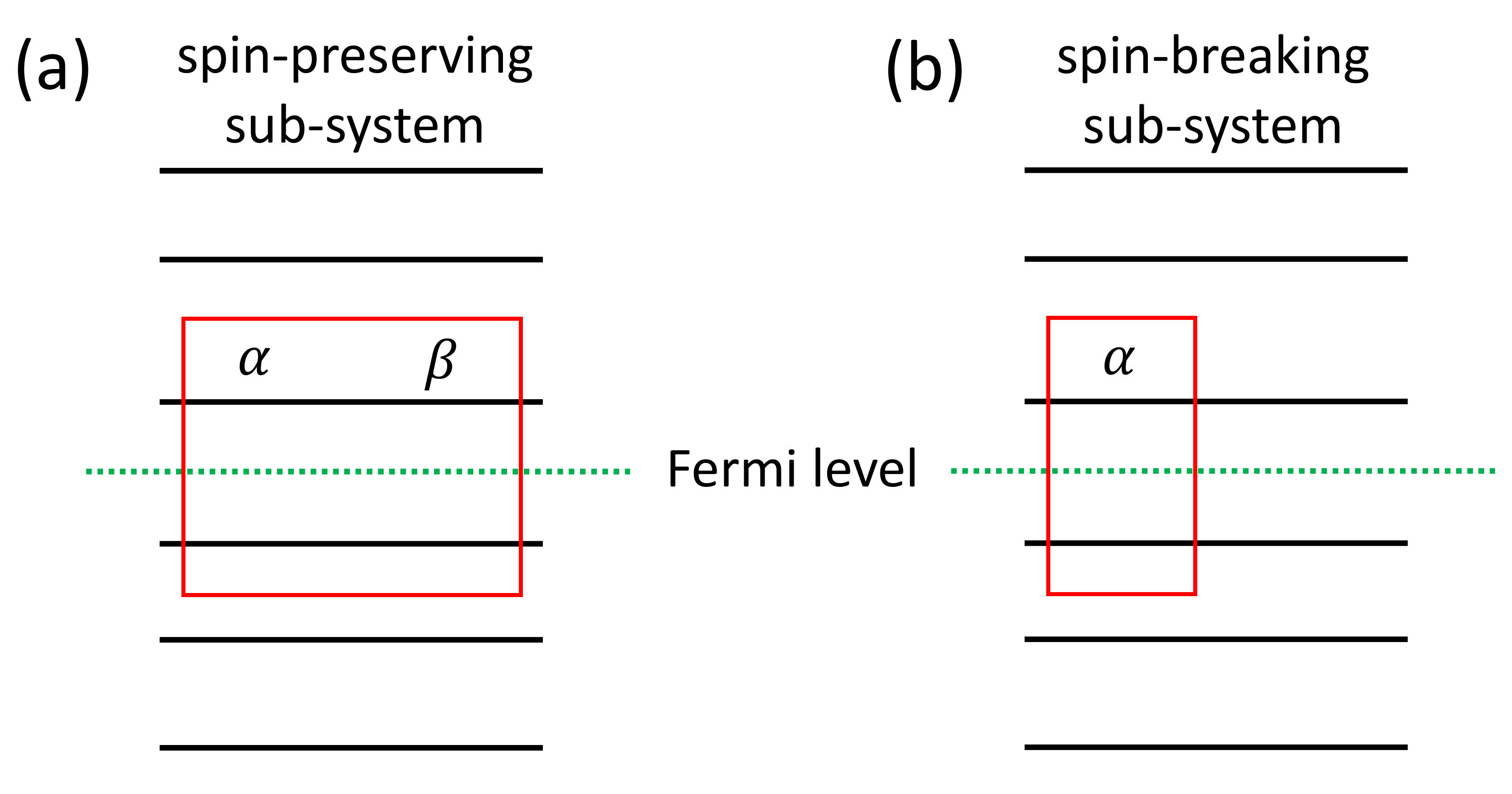}
    \caption{Two types of active spaces considered in this paper: (a) orbital definition with the sub-system wave function $|\Psi(\mathfrak{h})\rangle$ preserving the symmetry of the reference  function 
    $|\Phi\rangle$ (see Eq.(\ref{subwf}), and (b)  spin-orbital definition of the active space - in this case the symmetry (especially the spin symmetry)  of the sub-system may be broken.}
    \label{fig:siam_results}
\end{figure}
As mentioned earlier, each standard CC approximation is characterized by its own class of SESs (see Ref.\onlinecite{safkk} for details).

The SES-CC theorem is valid for arbitrary SES  $\mathfrak{h}$ corresponding to a given CC approximation. For example, for the CCSD approximation, 
arbitrary non-trivial (i.e., containing at least one active occupied and one virtual active orbitals)  SES $\mathfrak{g}^{(N)}(R,S)$ defined at the orbital level, contains either one occupied active orbital or one virtual active orbital. This means that one can form $S_{\rm CCSD}$,
\begin{equation}
S_{\rm CCSD} =n_o (2^{n_v}-1) + n_v (2^{n_o}-1) \;,
\label{sccsd}
\end{equation}
various SESs and corresponding effective Hamiltonians, that upon diagonalization,  reproduce the standard CCSD energy
(this formula is a consequence of binomial expansion, and  that $k$ active virtual/occupied orbitals can be chosen in 
${n_v \choose k}$/${n_o \choose k}$  different ways).
For higher-rank approximations, the number of SESs increases rapidly. 
In Eq.(\ref{sccsd}), $n_o$ and $n_v$ correspond to the number occupied and virtual orbitals. 
Since in the definition of the effective Hamiltonian, Eqs. (\ref{heffses}) and (\ref{heffdef}), only $T_{\rm ext}(\mathfrak{h})$ is involved, one can view the SES CC formalism with the resulting active-space eigenvalue problem, Eq. (\ref{seseqh}), as a specific form of a renormalization procedure where external parameters defining the corresponding wave function are integrated out. One should also mention that calculating the CC energy 
using Eq. (\ref{seseqh}), is valid for {\it any SES for a given   CC approximation defined  by  cluster operator $T$}.
In line with the SES-CC theorem, 
 the standard CC energy expression, shown by Eq. (\ref{coeq2}), can be reproduced when one uses a  trivial sub-algebra, which contains no excitations (i.e., active space is spanned by the  $|\Phi\rangle$ only).

The SES-CC theorem can be extended to the case when several SES-CC non-Hermitian eigenvalue problems are integrated into the so-called quantum flow 
(see Refs. \onlinecite{safkk,kowalski2021dimensionality,bauman2022coupled})
composed of coupled low-dimensionality eigenproblems
\begin{widetext}
\begin{equation}
    H^{\rm eff}(\mathfrak{h}_i)
    e^{T_{\rm int}(\mathfrak{h}_i)}|\Phi\rangle = 
    E e^{T_{\rm int}(\mathfrak{h}_i)}|\Phi\rangle  \;
    (i=1,\ldots,M_{\rm SES})\;,
\label{seseqhf}
\end{equation}
\end{widetext}
where $M_{\rm SES}$ is the total number of SESs or active space problems included in the flow. In Ref.\onlinecite{kowalski2021dimensionality,bauman2022coupled}, we demonstrated that problem defined in this way is equivalent (at the solution) to the standard CC equations given by Eqs.(\ref{coeq1}) and (\ref{coeq2}) with cluster operator defined as 
a combination of all {\em unique} excitations included in 
$T_{\rm int}(\mathfrak{h}_i)$ $(i=1,\ldots,M_{\rm SES})$ operators.
This observation allows to (1) capture the local correlation effects in a more controllable way (if all SESs involved in the quantum flow are defined by pairs of active occupied and localized orbitals, then each eigenvalue sub-problem is defined by the effective Hamiltonian, which inherently allows one to define pair density matrix; moreover each "pair" sub-problem corresponds to a correlation of fours electrons, which using the same density matrix enables one to select subsets of triple and quadruple excitations in additions to singles and doubles), and (2) re-represent the process of solving equations for many-body systems - instead of treating the whole system at once, 
one can deal with only one reduced-dimension
sub-system  at the time.
In quantum computing, the unitary variant of the SES-CC formalism \cite{kowalski2021dimensionality} can also be used to introduce constant-depth quantum algorithms.

It is also interesting to analyze the behavior of the CC equations in the context of the non-interacting sub-subsystem limit (NSL). Let us assume that we partition the entire system $\Delta$ into $K$ sub-systems
$\lbrace \Delta_i \rbrace_{i=1}^K$ and approximate cluster operator contains all components $T_{\Delta_i}$ needed to fully correlate individual sub-systems $\Delta_i$ in the NSL (for example, for CCSD theory sub-systems $\Delta_i$ are two electron systems). Therefore, in the NSL, the cluster operator $T$ and Hamiltonian $H$ can be written (we assume that localized basis set is employed) as sums of components $T_{\Delta_i}/H_{\Delta_i}$ describing sub-systems:
\begin{equation}
T = \sum_{i=1}^{K} T_{\Delta_i} \;\;,\;\;  
H = \sum_{i=1}^{K} H_{\Delta_i} \;,\; 
\label{nsl2}
\end{equation}
Given the commutativity relations between $H_{\Delta_i}$'s ($\lbrack H_{\Delta_i},H_{\Delta_j} \rbrack=0$) one can derive the following form of the CC equations in the NSL limit:
\begin{widetext}
\begin{equation} 
(P+Q_{\Delta_m})\lbrack e^{-\sum_{i\ne m} T_{\Delta_i}} (\sum_{j=1}^{K} H_{\Delta_j} ) e^{\sum_{k\ne m} T_{\Delta_k}}\rbrack e^{T_{\Delta_m}} |\Phi\rangle = E e^{T_{\Delta_m}} |\Phi\rangle 
\;,(m=1,\ldots,K)\;,
\label{nsl3}
\end{equation}
\end{widetext}
where $Q_{\Delta_m}$ is a projection operator onto excited Slater determinants localized on sub-system $\Delta_m$. It takes precisely the same form as equations (\ref{seseqhf}). 
Henceforth one can view quantum flow equations given by Eqs.(\ref{seseqhf}) 
as the extension of the properties of the CC equations in the NSL to the interaction regime, where all sub-systems interact.
On the other hand,  the form of CC equations in NSL is a special case of (\ref{seseqh}), where groups of active orbitals are at an infinite distance from each other. 

\section{Numerical Simulations} 

The numerical studies for several benchmark systems, including H4, H6, H8, and Li$_2$ systems, were performed using occupation-number-representation-based many-body manipulator code (stringMB)
that allows one to construct a matrix representation (${\bf A}$) of general second-quantized operators $A$. 
In particular, stringMB can be used to build matrix representations of the Hamiltonian, the external part of the cluster operator, and exponents of  $T_{\rm ext}(\mathfrak{h})$ for arbitrary $\mathfrak{h}$, i.e., 
\begin{eqnarray}
H &\rightarrow & {\bf H} \;\;, \label{hmat} \\
T_{\rm ext}(\mathfrak{h}) &\rightarrow & {\bf T}_{\rm ext}(\mathfrak{h}) \;\;, \label{textmat} \\
e^{-T_{\rm ext}(\mathfrak{h})}, \;  e^{T_{\rm ext}(\mathfrak{h})} &\rightarrow &  e^{-{\bf T}_{\rm ext}(\mathfrak{h})}, \;  e^{{\bf T}_{\rm ext}(\mathfrak{h})} \;. \label{expmat}
\end{eqnarray}
Moreover, the stringMB can extract sub-blocks  of matrices or their products corresponding to  arbitrary sub-space of the entire space. This feature is used to form matrix representations of the 
effective Hamiltonians $H^{\rm eff}(\mathfrak{h})$. 

We chose the STO-3G  and 6-31G (for beryllium atom only)  basis sets \cite{hehre1969self,pople72_2257} to describe benchmark systems considered here. To provide the numerical illustration of the SES-CC theorem in a variety of situations, for H4 and H6, we chose several geometries corresponding to weakly ($\alpha$=0.500 for H4 and $R_{\rm H-H}=2.0$ a.u. for H6 and H8) and strongly ($\alpha$=0.005 for H4 and $R_{\rm H-H}=3.0$ a.u. for H6) correlated ground states.


%
%
%
%

\section{CC theory and sub-subsystem consistency}

In this Section, using  the CCSD 
and CCSDTQ approaches
as  examples, we will illustrate that solving the CC equation is equivalent to establishing a self-consistency between various sub-systems defined by the corresponding SESs. Specifically, we will illustrate that the diagonalization of effective Hamiltonians corresponding to various SESs reproduces the CC energies obtained with Eq.(\ref{coeq2}). In Table \ref{tab:ground1} we collated ground-state eigenvalues $E(\mathfrak{h})$ of various SESs $\mathfrak{h}$ and corresponding effective Hamiltonians $H^{\rm eff}(\mathfrak{h})$ for H4 system in linear ($\alpha=0.500$) and almost square ($\alpha=0.005$) configurations and H6 for $R_{\rm H-H}$=2.0 a.u. and 
$R_{\rm H-H}$=3.0 a.u. For testing purposes for H4 we used the following CCSD SESs $\mathfrak{h}$: $\mathfrak{g}^{(N)}(1_R,1_S)$ with $R=\lbrace 2 \rbrace$, $S=\lbrace 3 \rbrace$, 
$R=\lbrace 1 \rbrace$, $S=\lbrace 3 \rbrace$, and $R=\lbrace 2 \rbrace$, $S=\lbrace 4 \rbrace$, $\mathfrak{g}^{(N)}(1_R,2_S)$ with $R=\lbrace 2 \rbrace$, $S=\lbrace 3,4 \rbrace$, and 
$\mathfrak{g}^{(N)}(2_R,1_S)$ with $R=\lbrace 1,2 \rbrace $, $S=\lbrace 3 \rbrace$. For each case considered in Table \ref{tab:ground1}, each value of $E(\mathfrak{h})$ reproduces exact value of the CCSD energy $E_{\rm CCSD}$. Analogous situations can be observed for the H6 system with 
the following SESs: 
$\mathfrak{g}^{(N)}(1_R,1_S)$ with $R=\lbrace 3 \rbrace$,  $S=\lbrace 4 \rbrace$, 
$\mathfrak{g}^{(N)}(1_R,1_S)$ with $R=\lbrace 2 \rbrace$,  $S=\lbrace 4 \rbrace$, 
$\mathfrak{g}^{(N)}(1_R,1_S)$ with $R=\lbrace 1 \rbrace$,  $S=\lbrace 6 \rbrace$, 
$\mathfrak{g}^{(N)}(1_R,2_S)$ with $R=\lbrace 1 \rbrace$,  $S=\lbrace 4,5 \rbrace$, and
$\mathfrak{g}^{(N)}(1_R,2_S)$ with $R=\lbrace 1 \rbrace$,  $S=\lbrace 4,5 \rbrace$
Again, for each case considered in Table \ref{tab:ground1}, each value of $E(\mathfrak{h})$ reproduces exact value of the CCSD energy $E_{\rm CCSD}$. 
\renewcommand{\tabcolsep}{0.15cm}
\begin{table*}[!ht]
    \centering
    \begin{tabular}{cccccc}
\hline\hline \\[-0.2cm]
\multicolumn{6}{c}{STO-3G H4 $\alpha$=0.005 $E_{\rm CCSD}$=-1.946325}\\
\hline \\[-0.2cm]
 &  $R$=$\lbrace 2\rbrace$, $S$=$\lbrace 3 \rbrace $ &
    $R$=$\lbrace 1\rbrace$, $S$=$\lbrace 3 \rbrace $ &
    $R$=$\lbrace 2\rbrace$, $S$=$\lbrace 4 \rbrace $ &
    $R$=$\lbrace 2\rbrace$, $S$=$\lbrace 3,4 \rbrace $ & 
    $R$=$\lbrace 1,2\rbrace$, $S$=$\lbrace 3 \rbrace $ \\[0.1cm]
\cline{2-6} \\
$E_{\mathfrak{h}}$ & 
-1.946325 & -1.946325 & -1.946325 & -1.946325 & -1.946325 \\[0.1cm]
\hline  \\[-0.2cm]
\multicolumn{6}{c}{STO-3G H4 $\alpha=0.500$ $E_{\rm CCSD}$=-2.151004}\\
\hline \\
 &  $R$=$\lbrace 2\rbrace$, $S$=$\lbrace 3 \rbrace $ &
    $R$=$\lbrace 1\rbrace$, $S$=$\lbrace 3 \rbrace $ &
    $R$=$\lbrace 2\rbrace$, $S$=$\lbrace 4 \rbrace $ &
    $R$=$\lbrace 2\rbrace$, $S$=$\lbrace 3,4 \rbrace $ & 
    $R$=$\lbrace 1,2\rbrace$, $S$=$\lbrace 3 \rbrace $ \\[0.1cm]
\cline{2-6} \\
$E_{\mathfrak{h}}$ & 
-2.151004  & -2.151004 & -2.151004 & -2.151004 & -2.151004 \\[0.1cm]
\hline  \\[-0.2cm]
\multicolumn{6}{c}{STO-3G H6 $R_{H-H}=2.0$ a.u. $E_{\rm CCSD}$=-3.217277}\\
\hline \\
 &  $R$=$\lbrace 3\rbrace$, $S$=$\lbrace 4 \rbrace $ &
    $R$=$\lbrace 2\rbrace$, $S$=$\lbrace 4 \rbrace $ &
    $R$=$\lbrace 1\rbrace$, $S$=$\lbrace 6 \rbrace $ &
    $R$=$\lbrace 1\rbrace$, $S$=$\lbrace 4,5 \rbrace $ & 
    $R$=$\lbrace 3\rbrace$, $S$=$\lbrace 4,5,6 \rbrace $ \\[0.1cm]
\cline{2-6} \\
$E_{\mathfrak{h}}$ & 
 -3.217277 & -3.217277 & -3.217277 & -3.217277 & -3.217277 \\[0.1cm]
\hline  \\[-0.2cm]
\multicolumn{6}{c}{STO-3G H6 $R_{H-H}=3.0$ a.u. $E_{\rm CCSD}$=-2.967326}\\
\hline \\
 &  $R$=$\lbrace 3\rbrace$, $S$=$\lbrace 4 \rbrace $ &
    $R$=$\lbrace 2\rbrace$, $S$=$\lbrace 4 \rbrace $ &
    $R$=$\lbrace 1\rbrace$, $S$=$\lbrace 6 \rbrace $ &
    $R$=$\lbrace 1\rbrace$, $S$=$\lbrace 4,5 \rbrace $ & 
    $R$=$\lbrace 3\rbrace$, $S$=$\lbrace 4,5,6 \rbrace $ \\[0.1cm]
\cline{2-6} \\
$E_{\mathfrak{h}}$ & 
 -2.967326 & -2.967326 & -2.967326 & -2.967326 & -2.967326 \\[0.1cm]
\hline  \\[-0.2cm]
    \end{tabular}
    \caption{Comparisons of the CCSD energies $E_{\rm CCSD}$ obtained with the standard textbook formula for CC energy with eigenvalues  $E_{\mathfrak{h}}$ of effective Hamiltonians acting in SES active spaces defined by various $R$ and $S$ orbital subsets. The tests were performed for H4  system defined by two values of geometry parameter $\alpha$ ($\alpha=0.005$ and $\alpha=0.500$) and H6 system, a linear chain of hydrogen atoms, for $R_{H-H}$=2.0a.u. and $R_{H-H}$=3.0a.u. All energies are reported in Hartree.}
    \label{tab:ground1}
\end{table*}
In Table \ref{tab:ground2} we collected $E(\mathfrak{h})$ for spin-orbital definitions of $\mathfrak{h}$ corresponding to the simplest spin-orbital-type active spaces where a single $\alpha$ electron is correlated 
within one occupied and one virtual active $\alpha$ spin-orbitals. 
As in the case of the orbital-type SESs, 
the $E(\mathfrak{h})$'s for H4, H6, and Li$_2$ reproduce exact values of $E_{\rm CCSD}$.
%
%
%
\renewcommand{\tabcolsep}{0.35cm}
\begin{table*}[!ht]
    \centering
    \begin{tabular}{cccc}
\hline\hline \\[-0.2cm]
 & \multicolumn{1}{c}{H4\;,\; $\alpha$=0.005} 
 & \multicolumn{1}{c}{H6\;,\; $R_{\rm H-H}$=2.0a.u.}
 & \multicolumn{1}{c}{Li$_2$\;,\; $R_{\rm Li-Li}$=2.673\AA} \\[0.2cm]
   \cline{2-4}  \\[0.0cm]
 & $E_{\rm CCSD}$=-1.946325 
 & $E_{\rm CCSD}$=-3.217277 
 & $E_{\rm CCSD}$=-14.667260 \\[0.2cm]
 & $R$=$\lbrace 1\alpha\rbrace$, $S$=$\lbrace 3\alpha \rbrace$
 & $R$=$\lbrace 3\alpha\rbrace$, $S$=$\lbrace 5\alpha \rbrace$
 & $R$=$\lbrace 1\alpha\rbrace$, $S$=$\lbrace 4\alpha \rbrace$ \\[0.2cm]
  \cline{2-4}  \\[0.1cm]
$E_{\mathfrak{h}}$ & -1.946325 & -3.217277 & -14.667260 
 \\[0.1cm]
\hline  \\[-0.2cm]
    \end{tabular}
    \caption{Comparisons of the CCSD energies $E_{\rm CCSD}$ obtained with the standard textbook formula for CC energy with eigenvalues  $E_{\mathfrak{h}}$ of effective Hamiltonians acting in SES active spaces defined by various $R$ and $S$ spin-orbital subsets. For example, $R$=$\lbrace 1\alpha\rbrace$, $S$=$\lbrace 3\alpha \rbrace$, designates spin-orbital active space spanned by $\alpha$ spin associated with orbital $1$ and $\alpha$ spin associated with orbital $3$. The $R_{\rm Li-Li}=2.673 \AA$  corresponds to near equilibrium geometry of the Li$_2$ molecule. All energies are reported in Hartree.}
    \label{tab:ground2}
\end{table*}
%
%
%
As a part of the discussion regarding the possibility of breaking symmetries of the whole quantum system, it is interesting to explore the possibility of breaking orbital energy degeneracies of the sub-system. To extend the previous paragraph's discussion, we chose the Be atom in the 6-31 basis set as a benchmark system. The degenerate RHF orbitals 3,4 and 5, as well as 7, 8, and 9, correspond to different $p$ shells. In our simulations, we used the following active spaces: (1)  $R$=$\lbrace 2 \rbrace$, $S$=$\lbrace 3\rbrace$, (2) 
$R$=$\lbrace 2 \rbrace$, $S$=$\lbrace 3,4  \rbrace$, 
(3) 
$R$=$\lbrace 2 \rbrace$, $S$=$\lbrace 3,4,5 \rbrace$, and 
(4) $R$=$\lbrace 2 \rbrace$, $S$=$\lbrace 3,9 \rbrace$. 
While the active space (3) would usually be involved in typical quantum chemical calculations, the active spaces (1), (2), and (4) correspond to the unusual situation where the degeneracies of the orbital energies are broken. The active space (4) also contains two active virtual orbitals corresponding to two separate $p$ shells. Despite this fact, as seen from Table \ref{tab:ground2a}, the SES-CC theorem produces energies equal to the CCSD energy in all situations. This is yet another illustration of an interesting feature of the SES-CC theorem associated with various scenarios for symmetry breaking. 
%
%
\renewcommand{\tabcolsep}{0.35cm}
\begin{table*}[!ht]
    \centering
    \begin{tabular}{ccccc}
\hline\hline \\[-0.2cm]
  & \multicolumn{4}{c}{Be atom} \\[0.2cm]
    \cline{2-5}  \\[0.0cm]
 & \multicolumn{4}{c}{$E_{\rm CCSD}$=-14.613518 }
  \\[0.2cm]
 & $R$=$\lbrace 2 \rbrace$, $S$=$\lbrace 3 \rbrace$
 & $R$=$\lbrace 2 \rbrace$, $S$=$\lbrace 3,4 \rbrace$
 & $R$=$\lbrace  2 \rbrace$, $S$=$\lbrace  3,4,5 \rbrace$ 
 & $R$=$\lbrace  2 \rbrace$, $S$=$\lbrace  3,9 \rbrace$ \\[0.2cm]
  \cline{2-5}  \\[0.1cm]
$E_{\mathfrak{h}}$ & -14.613518 & -14.613518 & -14.613518 & -14.613518
 \\[0.1cm]
\hline  \\[-0.2cm]
    \end{tabular}
    \caption{ Comparisons of the CCSD energies $E_{\rm CCSD}$ obtained with the standard textbook formula for CC energy with eigenvalues  $E_{\mathfrak{h}}$ of effective Hamiltonians acting in the SES generated  active spaces defined by various $R$ and $S$ orbital subsets for the Be atom  in the 6-31G basis set. All energies are reported in Hartree.}
    \label{tab:ground2a}
\end{table*}
\renewcommand{\tabcolsep}{0.35cm}
\begin{table*}[!ht]
    \centering
    \begin{tabular}{ccccc}
\hline\hline \\[-0.2cm]
  & \multicolumn{4}{c}{H6\;,\; $R_{\rm H-H}$=2.0a.u.} \\[0.2cm]
    \cline{2-5}  \\[-0.2cm]
 & \multicolumn{4}{c}{$E_{\rm CCSDTQ}$=-3.217699 }
  \\[0.2cm]
 & $R$=$\lbrace 2,3 \rbrace$, $S$=$\lbrace 4,5 \rbrace$
 & $R$=$\lbrace 1,2 \rbrace$, $S$=$\lbrace 5,6 \rbrace$
 & $R$=$\lbrace  3 \rbrace$, $S$=$\lbrace  4 \rbrace$ 
 & $R$=$\lbrace  3 \rbrace$, $S$=$\lbrace  4,5 \rbrace$ \\[0.2cm]
  \cline{2-5}  \\[0.1cm]
$E_{\mathfrak{h}}$ & -3.217699 & -3.217699 & -3.217699 & -3.217699
 \\[0.1cm]
\hline  \\[-0.2cm]
    \end{tabular}
    \caption{ Comparisons of the CCSDTQ energies $E_{\rm CCSDTQ}$ obtained with the standard textbook formula for CC energy with eigenvalues  $E_{\mathfrak{h}}$ of effective Hamiltonians acting in the SES generated  active spaces defined by various $R$ and $S$ orbital subsets for the H6 linear chain with $R_{H-H}=2.0$ a.u. in the STO-3G basis set. All energies are reported in Hartree.}
    \label{tab:ground3}
\end{table*}
\renewcommand{\tabcolsep}{0.35cm}
\begin{table*}[!ht]
    \centering
    \begin{tabular}{ccccc}
\hline\hline \\[-0.2cm]
  & \multicolumn{4}{c}{H8\;,\; $R_{\rm H-H}$=2.0a.u.} \\[0.2cm]
    \cline{2-5}  \\[-0.2cm]
 & \multicolumn{4}{c}{$E_{\rm CCSDTQ}$=-4.286013 }
  \\[0.2cm]
 & $R$=$\lbrace 3,4 \rbrace$, $S$=$\lbrace 5,6 \rbrace$
 & $R$=$\lbrace 1,2 \rbrace$, $S$=$\lbrace 7,8 \rbrace$
 & $R$=$\lbrace  4 \rbrace$, $S$=$\lbrace  5 \rbrace$ 
 & $R$=$\lbrace  4 \rbrace$, $S$=$\lbrace  5,6 \rbrace$ \\[0.2cm]
  \cline{2-5}  \\[0.1cm]
$E_{\mathfrak{h}}$ & -4.286013 & -4.286013& -4.286013 & -4.286013
 \\[0.1cm]
\hline  \\[-0.2cm]
    \end{tabular}
    \caption{Comparisons of the CCSDTQ energies $E_{\rm CCSDTQ}$ obtained with the standard textbook formula for CC energy with eigenvalues  $E_{\mathfrak{h}}$ of effective Hamiltonians acting in the SES generated  active spaces defined by various $R$ and $S$ orbital subsets for the H8 linear chain with $R_{H-H}=2.0$ a.u. in the STO-3G basis set. All energies are reported in Hartree.}
    \label{tab:ground4}
\end{table*}
%

As mentioned earlier, the SES-CC theorem is valid for standard CC approximation defined by a given excitation level. While our discussion has focused so far on the CCSD methods, in the following part of this Section, we will discuss the SES-CC theorem in the context of the CCSDTQ formulation. To this end, we performed a series of calculations for systems containing more than four correlated electrons where the CCSDTQ approach does not correspond to the exact theory. In Tables \ref{tab:ground3} and \ref{tab:ground4}, we collated the results of the SES-CC simulations for H6 and H8 models in the STO-3G basis set. 

For H6, we considered two types of SES-CC  active spaces containing two occupied and two virtual orbitals, i.e., $R$=$\lbrace 2,3 \rbrace$, $S$=$\lbrace 4,5 \rbrace$ and 
$R$=$\lbrace 1,2 \rbrace$, $S$=$\lbrace 5,6 \rbrace$. While the first active space is directly related to the corresponding ground-state problem, the second one contains excited Slater determinants 
(with respect to the reference function $|\Phi\rangle$) that represent a small contribution to the ground-state wave function. In both cases, the diagonalization of the corresponding effective 
Hamiltonians $H^{\rm eff}(\mathfrak{h})$ leads to the eigenvalues $E_{\mathfrak{h}}$  exactly reproducing the CCSDTQ energy $E_{\rm CCSDTQ}$. One should stress that the above active spaces are SES-CC spaces for the CCSDTQ level of the theory (the CCSDTQ ansatz based on the $T_{\rm int}(\mathfrak{h})$ operator generates FCI-type expansion in these active spaces) and not for the CCSD level of theory. However, as discussed in Ref.\onlinecite{safkk}, the SESs for higher-level theory contain all SESs for lower-rank approaches. To provide numerical confirmation of this statement, we performed calculations using CCSD-type SES, i.e., 
$R$=$\lbrace 3 \rbrace$, $S$=$\lbrace 4 \rbrace$ and 
$R$=$\lbrace 3 \rbrace$, $S$=$\lbrace 4,5 \rbrace$. Again, in both cases we were able to reproduce the CCSDTQ energies. 

Analogous analysis can be performed for the H8 system (see Table \ref{tab:ground4}). For all types of SES active spaces discussed in Table \ref{tab:ground4}, we reproduced the exact values of the CCSDTQ energy. 

For H6 and H8 systems, using the CCSDTQ formalism, we also performed  simulations based on the  spin-orbital definition of $\mathfrak{h}$, e.g., 
$R$=$\lbrace 2\alpha\rbrace$, $S$=$\lbrace 4\alpha \rbrace$ (H6) and 
$R$=$\lbrace 4\alpha\rbrace$, $S$=$\lbrace 6\alpha \rbrace$ (H8) that reproduce the exact values of the CCSDTQ energies using sub-systems composed of a single correlated electron in two active spin-orbitals. 
%
\section{Conclusions}
Our numerical tests indicate that CC energies can be obtained alternatively to the textbook energy formula by diagonalizing effective Hamiltonians for sub-systems defined in appropriate active spaces. In the present studies, we focused on the closed-shell CC  formulations where the  active spaces can be determined in terms of active orbitals or active spin-orbitals. We demonstrated that the CC energies for the CCSD and CCSDTQ approaches could be exactly reproduced by using these two types of active spaces. In the extreme case, we showed that the CC energy could be reproduced by a sub-system (in the sense of sub-system wave function defined by Eq.(\ref{subwf})) composed of one active electron in two active  $\alpha$-type spin-orbitals. We also demonstrated that the alternative ways of obtaining CC energy could be viewed as an analog to the asymptotic behavior of the CC formalism in the non-interacting sub-system limit in the presence of interactions. 
These facts have interesting consequences regarding how CC theory should be interpreted and how a new class of CC approximations can be formulated. The main conclusions are listed below: 
\begin{itemize}
\item From the SES-CC perspective, the standard single-reference CC Ansatz can be viewed as a renormalization procedure where energies describing all possible SES-CC sub-systems are calculated self-consistently.  
\item The quantum flow formalism introduced and discussed in Refs.\onlinecite{safkk,kowalski2021dimensionality} exploits this feature to define a new class of approximations that self-consistently correlate pre-defined classes of sub-systems. At the convergence, all ground-state eigenvalues of all effective Hamiltonians are equal and correspond to the approximate ground-state energy of the entire system.  
\item The size-consistency of approximations designed based on quantum flow equations is guaranteed by the so-called equivalence theorem (see Refs.\onlinecite{kowalski2021dimensionality,bauman2022coupled})
\item The quantum flow approach provides a natural way to capture the quantum system's sparsity. This is because each sub-system is described in terms of the corresponding non-Hermitian (Hermitian, for unitary CC flows)
eigenvalue problems defined by Hamiltonian $H^{\rm eff}(\mathfrak{h}_i)$ (see Eq.(\ref{seseqhf})), which enables one in a natural way to determine sub-system's one-body density matrix and select sub-system's natural orbitals to select class of the most important cluster amplitudes as in the local CC formulations \cite{neese2009efficient,neese2009efficient2,riplinger2013efficient,Neese16_024109}). 
\item 
The quantum flow equations represent a quantum many-body problem in terms of coupled,  reduced-dimensionality, and computable (both from the classical and quantum computing perspective)  sub-problems. 
It
can provide a new way of looking at the orthogonality catastrophe discussed in Ref.\cite{lee2022there}  (see also Refs.
\onlinecite{kohn1999nobel,chan2012low,mcclean2014exploiting,lee2022there}) both in the context of classical and quantum computing.
\end{itemize}
A part of the ongoing development is associated with the formulation of excited-state extensions of quantum flows that guarantee the size intensity of the quantum flows. The process assures this feature by picking up excited states localized on a chosen sub-system. 
This is especially important for unitary CC flows discussed in Ref.\onlinecite{kowalski2021dimensionality}, and excited-state applications of downfolding techniques in quantum computing.\cite{bauman2019quantumex}.




\section{Acknowledgement}
This  work  was  supported  by  the ``Embedding Quantum Computing into Many-body Frameworks for Strongly Correlated  Molecular and Materials Systems'' project, which is funded by the U.S. Department of Energy(DOE), Office of Science, Office of Basic Energy Sciences, the Division of Chemical Sciences, Geosciences, and Biosciences. 
All simulations were performed using Pacific Northwest National Laboratory (PNNL) computational resources. 
PNNL is operated for the U.S. Department of Energy by the Battelle Memorial Institute under Contract DE-AC06-76RLO-1830.

\section*{AUTHOR DECLARATIONS}
\subsection*{Conflict of Interest}
The author has no conflicts of interest to declare.

\section*{DATA AVAILABILITY}
The data that support the findings of this study are available from the corresponding authors
upon reasonable request.

%
%
%
\end{document}